\documentclass[12pt,a4paper]{article}

  \usepackage{a4wide}
  \usepackage{latexsym}
  \usepackage{epsf}
  \usepackage{amssymb}
  \usepackage{graphicx}
  \usepackage{amsmath, cite}
  \usepackage{slashed,epsfig}
  \usepackage{bbm}
  \usepackage{bbm}
  \usepackage{amsmath,amssymb,amsthm}
  \usepackage{pst-all}
  \newtheorem{lem}{Lemma}[section]

\renewcommand{\d}{\textrm{d}}
\newcommand{\e}{\textrm{e}}
\newcommand{\mf}[1]{\mathfrak{#1}}

\newcommand{\Real}{\textrm{I\!R}}
\newcommand{\mc}[1]{\mathcal{#1}}

\newcommand{\SU}{\mathop{\rm SU}}
\newcommand{\SO}{\mathop{\rm SO}}
\newcommand{\CSO}{\mathop{\rm CSO}}

\newcommand{\SL}{\mathop{\rm SL}}

\def\rme{{\mathrm e}}

\begin{document}

\begin{flushright}
\small
UG-06-07\\
KUL-TF-06-26\\
\date \\
\normalsize
\end{flushright}

\begin{center}

\vspace{.7cm}

{\LARGE {\bf Scaling Cosmologies of \\
\vspace{.7cm} $\mathcal{N}=8$
 Gauged Supergravity }} \\

\vspace{1.2cm}

\begin{center}
 Jan Rosseel$^{\dagger}$, Thomas Van Riet$^{\ddagger}$ and Dennis B. Westra$^{\ddagger}$ \\[3mm]
{\small\slshape
 $^{\dagger}$Institute for Theoretical Physics, K.U. Leuven,\\
Celestijnenlaan 200D, B-3001 Leuven, Belgium}\\
{\upshape\ttfamily jan.rosseel@fys.kuleuven.ac.be}\\[3mm]
\end{center}

   {\small\slshape
   $^{\ddagger}$Centre for Theoretical Physics, University of Groningen,\\
    Nijenborgh 4, 9747 AG Groningen, The Netherlands} \\
{\upshape\ttfamily t.van.riet,d.b.westra@rug.nl}\\[3mm]

\vspace{5mm}

{\bf Abstract}

\end{center}

\begin{quotation}

\small We construct exact cosmological scaling solutions in
$\mathcal{N}=8$ gauged supergravity. We restrict to solutions for
which the scalar fields trace out geodesic curves on the scalar
manifold. Under these restrictions it is shown that the axionic
scalars are necessarily constant. The potential is then a sum of
exponentials and has a very specific form that allows for scaling
solutions. The scaling solutions describe eternal accelerating and
decelerating power-law universes, which are unstable. An uplift of
the solutions to 11-dimensional supergravity is carried out and the
resulting timedependent geometries are discussed. In the discussion
we briefly comment on the fact that $\mathcal{N}=2$ gauged
supergravity allows accelerating stable scaling solutions.
\end{quotation}

\newpage

\pagestyle{plain} \tableofcontents
\section{Introduction}

In order to understand the possible (late-time) cosmological
scenarios in string theory it is natural to study this in a
supergravity context with a higher-dimensional origin
\cite{Kallosh:2001gr, VanProeyen:2006mf,
Kallosh:2002gf,Townsend:2003qv,Townsend:2001ea}. That way one can
learn how supersymmetry and a higher-dimensional origin constrain
the possibilities. In this light some investigations on the vacuum
structure of gauged extended supergravities have been carried out;
for instance de Sitter vacua in such theories can generically be
found
\cite{Kallosh:2001gr,Kallosh:2002gf,deRoo:2002jf,deRoo:2003rm,Gibbons:2001wy,Cvetic:2004km,Fre:2002pd,Cosemans:2005sj}.
However, it is only for $\mathcal{N}=2$ supergravity that stable de
Sitter vacua have been constructed
\cite{Fre:2002pd,Cosemans:2005sj}, unfortunately for those example
examples the higher-dimensional origin is unclear.

The possibilities for dark energy go beyond a positive cosmological
constant, see reference \cite{Copeland:2006wr} for a recent
overview. An interesting possibility is the existence of
cosmological scaling solutions. There are many definitions used but
a common feature is that scaling cosmologies correspond to
attractors, repellers and saddlepoints of the cosmological dynamical
system. The definition used here is that the ratio of the energy
densities of different constituents remains constant during
evolution. For a \emph{matter-scaling solution} the energy density
of the background barotropic fluid evolves in a constant ratio with
respect to the scalar field energy density. Since such solutions can
correspond to an attractor they are typically used in attempts to
alleviate the cosmic coincidence problem \cite{Amendola:2006qi}.
Scaling cosmologies are characterized by a scalefactor that is
power-law, $a(\tau)\sim \tau^P$. We refer to
\cite{Amendola:2006qi,Copeland:2006wr,Buchert:2006ya} for more
phenomenological issues concerning scaling solutions.

The scaling solutions studied in this paper are of two kinds, the
matter-scaling explained above and a \emph{scalar-dominated scaling
solution} in which the energy density of the barotropic fluid
vanishes and the potential energy of the scalar fields scales as the
kinetic energy\footnote{This is also true for matter-scaling
solutions so the only difference is that the fluid vanishes.}.

In contrast to de Sitter solutions scaling cosmologies have not been
given that much attention in supergravity. In reference
\cite{Townsend:2003qv} an unstable accelerating ($P>1$) scaling
solution with $P=3$ was found in $\mathcal{N}=8$ supergravity as an
alternative to acceleration from de Sitter solutions. In
\cite{deRoo:2006ms} an example of a stable scaling solution, with
$P=1$ was found in $\mathcal{N}=4$ gauged supergravity. Finally,
reference \cite{Tolley:2006ht} considered scaling solutions of
6-dimensional gauged chiral supergravity compactified to 4
dimensions.

It is the aim of this paper to systematically find the scaling
cosmologies in $\mathcal{N}=8$, $D=4$ supergravity and to check
their stability. Although $\mathcal{N}=8$ supergravity is not
realistic from a particle physics point of view it has attractive
features; there is only the supergravity multiplet and the different
theories only differ in the gauge group. This simplicity makes it
easier to oversee all the possibilities for finding interesting
solutions. It is believed that many (if not all) $\mathcal{N}=8$
theories have a higher-dimensional origin. For the gaugings we
consider in this paper the higher-dimensional origin is known
explicitly \cite{Hull:1988jw}.

To perform an exhaustive study of cosmological solutions in
supergravity theories is notoriously difficult because of the many
scalar fields and the corresponding complicated potentials. For
instance in $\mathcal{N}=8$ supergravity there are 70 scalars that
parametrize the E$_{7(+7)}/\SU(8)$-coset space and the complexity of
the potential depends on the gauge group of the theory. In this
respect, ungauged supergravity is easier since there is no
potential. If we restrict to FLRW-universes and redefine cosmic time
$\tau$ to a new time-coordinate $s$ via $\d \tau=a(\tau)^3\d s$, the
scalar field action in ungauged supergravities reads:
\begin{equation}
S_{\text{scalar}}=\int G_{ij}(\phi)\phi'^i\phi'^j ds\,,
\end{equation}
where $'$ indicates derivation with respect to $s$. This is the
action for geodesic curves parametrized by the affine parameter $s$.
Therefore cosmologies driven by massless fields are geodesics on the
scalar manifold. For the moduli spaces that appear in maximal
supergravity there have been investigations on the geodesic curves
and their higher-dimensional interpretation
\cite{Fre:2003ep,Fre:2005bs}.

In the case that there is a scalar potential reference
\cite{Karthauser:2006ix} stated that the scalars of all scaling
solutions describe geodesic curves on the scalar manifold. However
the proof of \cite{Karthauser:2006ix} shows that a geodesic can
indeed give rise to a scaling solution but the converse statement --
a scaling solution must be a geodesic -- is not proven. An example
of a scaling solution that does not describe a geodesic can be found
for the axion-dilaton system of reference \cite{Sonner:2006yn}. We
consider those scaling solutions that are geodesic as an interesting
subclass. Since geodesics on symmetric spaces are well understood we
can perform a systematic search for geodesic scaling cosmologies of
gauged supergravities. Ideally we would like to study all possible
scaling cosmologies and not just the class that describes geodesics.

We make some important restrictions in this paper. We consider only
flat FLRW-universes and ignore scaling solutions that exist on the
boundary of the scalar manifold\footnote{They are called
\emph{non-proper} solutions in references
\cite{Collinucci:2004iw,Hartong:2006rt}.}. We refer to
\cite{Hartong:2006rt} for a general treatment of scaling solutions
in the presence of spatial curvature, that also includes solutions
on the boundary of the scalar manifold.

\section{Scaling solutions and geodesics}

The action we consider contains $N$ Klein--Gordon fields $\phi^i$
coupled to gravity:
\begin{equation}
S=\int {\d}^4x\,\sqrt{-g}\Bigl[\frac{1}{2\kappa^2
}\mathcal{R}-\tfrac{1}{2}G_{ij}(\phi)\partial_{\mu}\phi^i\,\partial^{\mu}\phi^j-V(\phi)\Bigr]
 + S_{\text{matter}} \, ,
\end{equation}
where $\kappa^2=8\pi G$ with $G$ Newton's constant and
$S_{\text{matter}}$ is the action that describes a barotropic fluid
with constant equation-of-state-parameter $\gamma-1$. In a flat
FLRW-background, $\d s^2=-\d \tau^2+a(\tau)^2\d\vec{x}^2$, the
equations of motion read\footnote{We use the mostly plus convention
for the metric and the Riemann tensor reads
$\mathcal{R}^{\mu}_{\,\nu\rho\sigma}=\partial_{\rho}\Gamma_{\nu\sigma}^{\mu}+\ldots$}:
\begin{align}
& \ddot{\phi}^i+\Gamma^i_{jk}\dot{\phi}^j\dot{\phi}^k+3H\dot{\phi}^i=-G^{ij}\partial_j V\,, \\
& \dot{\rho}+3\gamma H \rho=0\,,\\
& p=(\gamma-1)\rho\,,\\
& H^2=\frac{\kappa^2}{3}(T+V+\rho) \,, \label{friedmann}\\
& \dot{H}=-\kappa^2 (T+\tfrac{1}{2}\gamma\rho)
\,,\label{acceleration}
\end{align}
with $T$ the kinetic energy of the scalars,
$T=\frac{1}{2}G_{ij}(\phi)\dot{\phi}^i\,\dot{\phi}^j$ and
$H=\dot{a}/a$, the Hubble parameter. We choose units in which
$\kappa^2=\tfrac{1}{2}$.

We define a scaling cosmology as a solution of the above equations
for which:
\begin{equation}\label{scaling}
V(\tau) \sim T(\tau) \sim \rho(\tau)\,.
\end{equation}
From the Friedmann equation (\ref{friedmann}) and the acceleration
equation (\ref{acceleration}) we find that the scalefactor is
power-law $a\sim \tau^P$ and vice versa. Hence for a scaling
solution we have:
\begin{equation}\label{scaling2}
V(\tau) \sim T(\tau) \sim \rho(\tau) \sim H^2(\tau) \sim
\dot{H}(\tau) \sim \frac{1}{\tau^2}\,.
\end{equation}

Scaling solutions for multiple fields are well studied for flat
scalar manifolds $G_{ij}=\delta_{ij}$ with (multiple) exponential
potentials
\cite{Liddle:1998jc,Malik:1998gy,Coley:1999mj,Copeland:1999cs,Collinucci:2004iw,Hartong:2006rt},
where the scaling solutions all have the following form:
\begin{equation}
\phi^i=a^i \ln(\tau)+ b^i\,.
\end{equation}
We notice that the scalars trace out straight lines, that is
geodesics on a flat space. If we take $s = \ln \tau$ as a parameter
the straight lines have a constant velocity $[\sum_{i}a_i^2]^{1/2}$
and the parameter $s$ is an affine parameter. The reason that
geodesics can describe scaling cosmologies comes from the constraint
$V\sim T$, which implies
\begin{equation}
\mathcal{L}_{\text{scalar}}\sim T(\tau)-V(\tau) \sim T(\tau)\,.
\end{equation}
Since $T=\tfrac{1}{2}G_{ij}\dot{\phi^i}\dot{\phi^j}\sim \tau^{-2}$
for scaling it is clear that $s=\ln \tau$ is an affine parameter
since $G_{ij}\partial_s\phi^i\partial_s\phi^j=\text{const}$. The
geodesic equation is:
\begin{equation}
\phi''^i+\Gamma^i_{jk}\phi'^j\phi'^k=0\,,
\end{equation}
where a prime denotes differentiation with respect to $s$.
Consistency with the Klein--Gordon equation gives rise to a first
order equation\cite{Karthauser:2006ix}:
\begin{equation}
(3P-1)\,\phi'^i=-e^{2s}G^{ij}\partial_j\, \ln V\,.
\end{equation}

\section{The $\mathcal{N}=8$ scalar potential}

The action of ungauged $\mathcal{N}=8$ supergravity in 4 dimensions
exhibits a rigid $\SL(8,\Real)$-symmetry \cite{Cremmer:1979up}. The
equations of motion allow for a larger, non-compact
E$_{7(+7)}$-symmetry. The theory contains 70 scalars that
parametrize the coset E$_{7(+7)}/\SU(8)$. De Wit and Nicolai gauged
the $\SO(8)$-subgroup of $\SL(8,\Real)$, by introducing minimal
couplings that break supersymmetry \cite{deWit:1982ig}. In order to
restore supersymmetry, further terms have to be added to the action.
In this way, one generates a potential that is proportional to the
square of the gauge coupling constant. Later on, starting from this
example, the so-called $\CSO(p,q,r)$-gaugings were found
\cite{Hull:1984rt,Hull:1984ea,Hull:1984wa}. The generators of this
subgroup are denoted by $\Lambda_{ab} = -\Lambda_{ba}$, with
$a,b=1,\cdots,8$ and they obey the following algebra:
\begin{equation} \label{algCSO}
[\Lambda_{ab}, \Lambda_{cd}] = \Lambda_{ad} \eta_{bc} - \Lambda_{ac}
\eta_{bd} - \Lambda_{bd} \eta_{ac} + \Lambda_{bc} \eta_{ad} \,,
\end{equation}
where
\begin{equation} \label{eta}
\eta_{ab} = \left( \begin{array}{ccc} \mathbbm{1}_{p \times p} & 0 & 0 \\
0 & - \mathbbm{1}_{q \times q} & 0 \\ 0 & 0 & 0_{r \times r}
\end{array} \right) \,.
\end{equation}
Other gaugings have been found later, see for instance
\cite{deWit:2003hr} but for the rest of the paper we only consider
the $\CSO$-gaugings.

Obtaining an explicit expression for the potential that can be used
to search for vacua, is a difficult task. Often one makes
truncations to get manageable expressions. Therefore we focus on the
$\SL(8,\mathbb{R})/\SO(8)$-submanifold of E$_{7(+7)}/\SU(8)$ that
contains the 35 scalar fields while the other 35 pseudoscalar fields
are consistently truncated. The action for the metric and the 35
scalars is given by \cite{Roest:2004pk}:
\begin{equation}
\mathcal{L}=\sqrt{-g}\Bigl\{\mathcal{R}
+\tfrac{1}{4}\text{Tr}[\partial\mathcal{M}\partial
\mathcal{M}^{-1}]-V \Bigr\}\,,
\end{equation}
where $\mathcal{M}=LL^T$ with $L$ the coset representative of the
$\SL(8,\mathbb{R})/\SO(8)$-coset. The potential is given by:
\begin{equation}\label{grouppotential}
V=\text{Tr}[(\eta\mathcal{M})^2]-\tfrac{1}{2}(\text{Tr}[\eta\mathcal{M}])^2\,.
\end{equation}
The scalar field equations of motion derived from the langrangian
are:
\begin{equation}
\partial[ \mathcal{M}^{-1}\partial \mathcal{M}]=4(\eta \mathcal{M})^2-2\text{Tr}[\eta \mathcal{M}]\eta \mathcal{M}-\frac{4}{n}V
\mathbbm{1}\,,
\end{equation}
with $n$ the dimension of the matrices (for now $n=8$).

The coset $\SL(8,\mathbb{R})/\SO(8)$ contains $7$ dilatons $\phi^i$
and $28$ axions $\chi^{\alpha}$. In the solvable gauge the coset
representative $L$ is written as
\begin{equation}
L=e^{\chi^{\alpha}E_{\alpha}}e^{-\tfrac{1}{2}\phi^i H_i}\,,
\end{equation}
where the sum over the indices $\alpha$ is a sum over the positive
root generators and the sum over the indices $i$ is a sum over the
Cartan generators .

For what follows we need some properties of the weights
$\vec{\beta}_a$ of the $\SL(n,\mathbb{R})$-algebra in the
fundamental representation:
\begin{equation}
\sum_a\beta_{ai}=0\,,\quad
\sum_a\beta_{ai}\beta_{aj}=2\delta_{ij}\,,\quad
\vec{\beta}_a\cdot\vec{\beta}_b=2\delta_{ab}-\frac{2}{n}\,.
\end{equation}
The last two identities hold in a handy basis that is given in
appendix \ref{appendix1}.

Because the axions appear at least squared in the potential it is
consistent to put them to zero\footnote{The kinetic term allows a
truncation of all the axions since the dilatons parameterize a
geodesic complete submanifold.}. Then the matrix $\mathcal{M}$
simplifies to
$\mathcal{M}=\text{diag}(e^{-\vec{\beta}_a\cdot\vec{\phi}})$, such
that the kinetic term becomes canonical and the potential is a sum
of exponentials:
\begin{equation}\label{potentiaal}
\mathcal{L}_{\text{scalar}}=-\tfrac{1}{2}\delta_{ij}\,\partial\phi^i\partial\phi^j-\tfrac{1}{2}\sum_{a=1}^{p+q}
e^{-2\vec{\beta}_a\cdot\vec{\phi}} +
\sum_{a<b}^{p+q}\eta_{aa}\eta_{bb}
e^{-(\vec{\beta}_a+\vec{\beta}_b)\cdot\vec{\phi}}\,.
\end{equation}

\section{Dilatonic scaling cosmologies}

The potential in (\ref{potentiaal}) is an example of a general
exponential potential, which is a sum of $M$ exponential terms that
depend on $N$ scalar fields:
\begin{equation}
V(\phi)=\sum_{a=1}^M\Lambda_a\,\text{exp}[\vec{\alpha}_a\cdot\vec{\phi}\,],
\end{equation}
where
$\vec{\alpha}_a\cdot\vec{\phi}=\sum_{i=1}^{N}\alpha_{ai}\phi^i$.
Scaling solutions of such potentials have been studied in great
detail
\cite{Copeland:1997et,Liddle:1998jc,Coley:1999mj,Malik:1998gy,Copeland:1999cs,Collinucci:2004iw,Hartong:2006rt}
and for convenience we reformulate some essential properties.

\subsubsection*{Scaling for multiple exponential potentials}

It follows from the autonomous system approach in
\cite{Collinucci:2004iw} that it is convenient to separate
exponential potentials in two classes I and II. Class I is
characterized by the fact that the $\vec{\alpha}_a$-vectors are
linearly independent whereas for class II they are linearly
dependent. Models that belong to the first class are known in the
literature under the name \emph{Generalized assisted inflation}
\cite{Copeland:1999cs}. Class I generically allows exact scaling
solutions, whereas class II can have exact scaling solutions only
when the $\vec{\alpha}_a$-vectors are \emph{affinely} related
\cite{Collinucci:2004iw,Hartong:2006rt}. Affinely related means that
there exists a set of $R$ independent $\vec{\alpha}_a$ such that
after relabelling $a=1\ldots R$, the remaining $\vec{\alpha}_{b}$
are expressed as $\vec{\alpha}_{b}=\sum_{a=1}^R
c_{ba}\vec{\alpha}_a$, where the coefficients $c_{ab}$ fulfill the
constraint:
\begin{equation}
\sum_{a=1}^R c_{ba}=1 \,,\quad\text{for all}\,\, b=R,\ldots, M\,.
\end{equation}
Both types of potentials that allow for scaling solutions have the
unique property that after an orthogonal field redefinition
$\vec{\phi}\rightarrow \vec{\varphi}$ the potential can always be
written as the following product \cite{Malik:1998gy,Hartong:2006rt}:
\begin{equation}\label{new potential}
V(\varphi)=e^{c\varphi_1}\,U(\varphi_2,\ldots,\varphi_{N})\,.
\end{equation}
Let us proof (\ref{new potential}) for class I and then for class II
with affinely related $\vec{\alpha}_a$-vectors. For the proof we
always assume that a field rotation is performed such that the
minimal number, $R$, of scalars appears in the potential and that
consequently $N-R$ scalar fields are free. This number $R$ equals
the number of linearly independent $\vec{\alpha}_a$-vectors
\cite{Collinucci:2004iw}. So class I has $R=M$ and class II $R<M$.

If the $\vec{\alpha}_a$ are linearly independent there exists a
(unit) vector $\vec{E}$ such that
\begin{equation}\label{innerproduct}
\vec{\alpha}_a\cdot\vec{E}=c\,,
\end{equation}
where $c$ is a number which is independent of the index $a$. Since,
if we multiply (\ref{innerproduct}) with $\alpha_{aj}$ and sum over
$a$ we obtain
\begin{equation}
\sum_{ai}\alpha_{aj}\,\alpha_{ai}\,E^i=c\sum_a\,\alpha_{aj}\,.
\end{equation}
The matrix $B_{ij}=\sum_a\alpha_{aj}\,\alpha_{ai}$ has an inverse
(because $R=M$) and the equation can be solved to find $E^i$. If we
now write the scalar fields in a different basis:
\begin{equation}
\vec{\phi}=\varphi_1\,\vec{E} +\vec{\varphi}_{\bot}\,,
\end{equation}
then we have in the new basis that $\alpha_{a1}=c$ for all $a$ and
consequently the potential takes the form (\ref{new potential}).

Now assume the $\vec{\alpha}_a$ are linearly dependent in an affine
way. Consider the $R$ independent vectors $\vec{\alpha}_{a}$ with
$a=1\ldots R$. For this subset we can repeat the same procedure as
above to find a unit vector $\vec{E}$ that obeys
(\ref{innerproduct}). Then we have in the new basis that
$\alpha_{a1}=c$ for $a=1\ldots R$. Consider $\alpha_{b1}$ for $b>R$:
\begin{equation}
\alpha_{b1}=\sum_{a=1}^R c_{ba}\alpha_{a1}=c\,\sum_{a=1}^R
c_{ba}=c\,.
\end{equation}
Again the potential can be factorized as in (\ref{new potential}).

It is easy to prove the inverse, if the potential can be written as
(\ref{new potential}) then either the $\vec{\alpha}_a$ are linearly
independent or they are dependent in an affine way.

As proven in \cite{Malik:1998gy,Hartong:2006rt} the exact scaling
solution is such that it is the overall scalar $\varphi_1$ that is
non-constant and the other scalars are constant. \emph{Therefore the
exact scaling solutions of multiple exponential potentials are such
that the potential is truncated to a single exponential
potential.\footnote{In \cite{Green:1999vv} the same was proven for
purely positive exponential terms and a special class of dilaton
couplings.}} The requirement for such a truncation is twofold.
Firstly, it must be possible to rewrite the potential like in
(\ref{new potential}) and secondly the function $U$ must have
stationary points ($\partial U =0$) in order to have a truncation
consistent with the equations of motion. If this is satisfied the
truncated action is given by:
\begin{equation}
S=\int \sqrt{-g}\Bigl[ \mathcal{R} -
\tfrac{1}{2}(\partial\varphi)^2-\Lambda\,e^{c\varphi}
\Bigr]+(S_{\text{Matter}})\,,
\end{equation}
where $\Lambda$ is the function $U$ at the stationary point. If the
scaling solution exists it is given by:
\begin{equation}\label{expliciet}
a\sim \tau^P\,, \quad \varphi=-\frac{2}{c}\ln
\tau+\frac{\ln(\frac{6P-2}{c^2\,\Lambda})}{c}\,,\quad
\rho=6(1-\frac{1}{c^2P})\frac{P^2}{\tau^2}.
\end{equation}
\begin{itemize}
\item Let us first assume that the barotropic fluid vanishes, then
the scaling solution is the scalar-dominated solution with
$P=1/c^2$. The scaling solution exists when $\Lambda>0$ and $P>1/3$
or $\Lambda<0$ and $P<1/3$. An inflationary solution ($P>1$)
requires $c^2<1$. The scaling solution with $\Lambda<0$ is never
stable and the scaling solution with $\Lambda>0$ is stable if the
extremum of $U$ is a minimum and the fluid perturbations imply an
extra stability condition $\tfrac{1}{c^2}>\tfrac{2}{3\gamma}$.

\item If on the other hand a barotropic fluid is nonzero there
exists a matter-scaling solution \cite{Copeland:1997et}. The
matter-scaling solution is such that the energy density of the
barotropic fluid and the scalar fields are non vanishing and have a
fixed ratio. The scalefactor of a matter-scaling solution is that of
a universe containing only the barotropic fluid, that is
$a\sim\tau^P$ with $P_{matter}=\tfrac{2}{3\gamma}$. The solution
exists when $\Lambda>0$ and $\tfrac{1}{c^2}<\tfrac{2}{3\gamma}$.
When $U$ is in a minimum the solution is stable.
\end{itemize}

\subsubsection*{The $\CSO$-dilaton potentials}

For the $\CSO(p,q,r)$-gaugings ($p+q+r=8$) with $r>0$ the potential
(\ref{grouppotential}) can be written in such a way that a smaller
coset matrix $\tilde{\mathcal{M}}$ appears. The result is
\cite{Roest:2004pk}:
\begin{equation}
V=e^{c\varphi}U(\tilde{\mathcal{M}})=e^{c\varphi}\Bigl[\text{Tr}[(\tilde{\eta}\tilde{\mathcal{M}})^2]-\tfrac{1}{2}(\text{Tr}[\tilde{\eta}\tilde{\mathcal{M}}])^2\Bigr]\,,
\end{equation}
where
\begin{equation}
c^2=\frac{8}{p+q}-1,\quad
\tilde{\eta}=\text{diag}(\mathbbm{1}_p,-\mathbbm{1}_q)\,,
\end{equation}
and the scalars appearing in $\tilde{\mathcal{M}}$ are those coming
from the $\frac{\SL(p+q,\Real)}{\SO(p+q)}$-coset and together with
$\varphi$ they span the manifold GL$(p+q,\Real)/\SO(p+q)=\Real\times
\SL(p+q,\Real)/\SO(p+q)$.

If we restrict to the dilatons (\ref{potentiaal}) becomes:
\begin{equation}
V=e^{c\varphi}U(\phi)=e^{c\varphi}\Bigl[\tfrac{1}{2}\sum
e^{-2\vec{\beta}_a\cdot\vec{\phi}} -
\sum_{a<b}\tilde{\eta}_{aa}\tilde{\eta}_{bb}
e^{-(\vec{\beta}_a+\vec{\beta}_b)\cdot\vec{\phi}}\Bigr]\,,
\end{equation}
where the vectors $\vec{\beta}_a$ are the weights of
SL$(p+q,\Real)$. For $c \neq 0$ this potential clearly belongs to
class II with affinely related $\vec{\alpha}_a$-vectors. Therefore
the potential is of the appropriate form for scaling solutions!

To find a scaling solution it is sufficient to find a stationary
point of $U$ that has the correct sign to allow for a scaling
solution. The stationary points of $U$ are most easily found using
lagrange multipliers as was shown in \cite{Cvetic:2004km}. The
outcome of this calculation is summarized in the table below.

\begin{table}[ht]
\begin{center}
\hspace{-1cm}
\begin{tabular}{|l|c|c|}
\hline \rule[-1mm]{0mm}{6mm} Gauging & matter-scaling $\rho_{\gamma}>0$ & scalar-dominated $\rho_{\gamma}=0$ \\
\hline \hline
1. $\CSO(1,0,7)$ & $P=2/(3\gamma)$& $\nexists$ \\
2. $\CSO(1,1,6)$ & $P=2/(3\gamma)$& $\nexists$ \\
3. $\CSO(2,2,4)$ & $\nexists$ & $P=1$ \\
4. $\CSO(3,3,2)$ & $\nexists$ & $P=3$ \\
5. $\CSO(4,3,1)$ & $\nexists$ & $P=7$ \\
\hline
\end{tabular}
\caption{\it $\mathcal{N}=8$ gaugings and their scaling solutions.}
\end{center}
\end{table}

We remark that there is no scalar-dominated scaling solution between
$1/3<P_{\text{scalar}}<1$. This implies that in these models a
matter-scaling solution can never coexist with a scalar-dominated
scaling cosmology.

The accelerating scaling solutions ($P>1$) are found for the
$\CSO(3,3,2)$-gauging and the $\CSO(4,3,1)$-gauging. The first was
found by Townsend in \cite{Townsend:2001ea} where it was constructed
by a reduction of a de Sitter vacuum in 5-dimensional
$\SO(3,3)$-gauged supergravity. The second possibility with $P=7$ is
as far as we know not found before. The cosmologies of the
$\CSO(1,1,6)$-gauging were considered before
\cite{Bergshoeff:2003vb} where the solutions were obtained from a
reduction of seven-dimensional pure gravity on a group manifold.

Since the matrix $\partial_i\partial_j U$ evaluated at an extremum
is not positive definite, the solutions are unstable.

\section{Axion-dilaton scaling cosmologies}

Once the axions are turned on the system is much more complex and
the construction of solutions becomes a difficult task. But if we
restrict to scaling cosmologies that describe geodesics the problem
boils down to parametrizing the geodesics and to check when they are
solutions. In what follows we describe a way to parametrize the
geodesics in terms of isometry transformations of straight lines.

Since the submanifold spanned by the dilaton fields is flat, the
geodesics on that part are straight lines
\begin{equation}
\phi^i(s)=v^i s+\phi^i(0)\,,
\end{equation}
in terms of the affine parameter introduced in section 2. Since
$\SL(n,\Real)$ is a symmetry of the geodesic equations it maps
geodesics to geodesics. Transforming the above straight line
generates a general geodesic, which is not a straight line. The
following lemma shows that in this way all geodesics can be
obtained. The proof is left for the appendix.

\begin{lem}
Every geodesic on the symmetric space $SL(n;\Real)/SO(n)$ can be
obtained by acting with isometries on a straight geodesic through
the origin.
\end{lem}

An isometry transformation is non-linear on the level at the
coordinates $\phi^i, \chi^{\alpha}$ but for the scalar matrix
$\mathcal{M}$ it works as:
\begin{equation}
\mathcal{M}(s) \rightarrow \Omega \mathcal{M}(s) \Omega^T,
\quad\quad \text{det}\Omega=1\,.
\end{equation}

The consequence of the above lemma is that the problem of finding
geodesic scaling solutions reduces to an algebraic problem. If a
geodesic scaling solution exists the scalars take the form:
\begin{equation}
\varphi=v_0 s+d_0\,,\quad \tilde{\mathcal{M}}=\Omega
\,D\,\Omega^T\,,
\end{equation}
where $D=\text{diag}(e^{-\vec{\beta}_i\cdot\vec{\phi}})$ with
$\vec{\phi}=\vec{v}s$. For simplicity we work in the truncated
system defined by $\tilde{\mathcal{M}}$ and $\varphi$, as explained
in the previous section.

For the scaling solutions we have
\begin{equation}
V(s)=\alpha\,e^{-2s} ,\quad H=P\,e^{-s}\,.
\end{equation}
With $\alpha$ some constant. If we substitute this in the equations
of motion for $\tilde{\mathcal{M}}$ and $\varphi$, we find the
following matrix equation:
\begin{equation}\label{matrix}
e^{-(2+cv_0)s-cd_0}\,\,A^{-1}\,\Bigl\{(1-3P)D^{-1}D'+\frac{4\alpha}{n}\mathbbm{1}\Bigr\}=4DAD-2\text{Tr}(AD)D\,,
\end{equation}
where $ A=\Omega^t\eta\Omega$. For a given matrix $D$ that
corresponds to a straight line, we can always make an orthogonal
field redefinition on the dilatons such that the matrix $D$
simplifies to\footnote{This is not an isometry transformation on the
coset, but corresponds to choosing a new coordinate system on the
coset. This new coordinate system does not affect equation
(\ref{matrix}).}:
\begin{equation}
D=\text{diag}(e^{-||\vec{v}||s},e^{+||\vec{v}||s},1,\ldots,1)\,.
\end{equation}
It is then not too difficult to check that the solutions of equation
(\ref{matrix}) necessarily have $||\vec{v}||=0$. This implies that
$\tilde{\mathcal{M}}$ is constant and the only running field is the
overall dilaton $\varphi$. In particular, if we act with the rigid
$\CSO$-symmetry on the dilaton solutions we are guaranteed to find
new solutions but only with constant axions.

\section{Higher-dimensional origin}

In \cite{Hull:1988jw} it was shown that the non-compact gaugings are
associated with 11-dimensional supergravity solutions that have a
non-compact internal space. For the $\CSO(p,q,r)$-gaugings the
internal space $\mathcal{H}^{p,q,r}$ is a hypersurface in
$\Real^{8}$, defined by the following equation:
\begin{equation} \label{hypersurface}
T_{AB} z^A z^B = R^2 \,\,\,\,\,\text{and}\,\,\,\,\,\,T=L^T \eta L\,,
\end{equation}
where $z^A$ are Cartesian coordinates of $\Real^{8}$ and $R$ is
determined by the flux of the 4-form field strength in 11
dimensions\footnote{The flux parameter $R$ is also the inverse of
the gauge coupling constant.}. In the previous sections we fixed the
flux parameter $R=1$. For arbitrary flux parameter the potential has
an extra factor $R^{-2}$ in front of the expression
(\ref{grouppotential}).

The metric on the hypersurface is then induced from the Euclidean
metric on $\Real^{8}$. Given a solution in 4 dimensions with a
metric $g_{4}$ and some scalars as only non-vanishing fields, the
11-dimensional metric $g_{11}$ is then determined as
\begin{equation} \label{11dmetric}
g_{11} = \Delta^{\frac{2}{3}} g_4(x) + \Delta^{-\frac{1}{3}}
g_{\mathcal{H}}(x,y)\,,
\end{equation}
where $g_{\mathcal{H}}(x,y)$ is the metric on $\mathcal{H}^{p,q,r}$
and $\Delta (x,y)$ is a warp factor
\begin{equation}
\Delta=\frac{T^2_{AB}z^Az^B}{R^2}\,.
\end{equation}

From the explicit solutions for the scalar matrix $\mathcal{M}$ we
notice that our scaling solutions correspond to $\SO(p)\times
\SO(q)$-invariant directions in the scalar coset\footnote{In terms
of the $\SO(p)\times\SO(q)$ invariant scalars s and t defined in
\cite{Kallosh:2001gr,Ahn:2001by,Gibbons:2001wy}, our solutions have
constant s and running t.}
\begin{equation} \label{etamatrix}
T = \rme^{\frac{c\varphi}{2}} \left( \begin{array}{ccc}
X\mathbbm{1}_{p \times p} & 0 & 0 \\
0 & - \bar{X}\mathbbm{1}_{q \times q} & 0 \\
0 & 0 & 0_{r \times r} \end{array} \right).
\end{equation}
The constants $X$ and $\bar{X}$ are the constant diagonal components
of the $\SL(p+q,\Real)/\SO(p+q)$-scalar matrix
$\tilde{\mathcal{M}}=\text{diag}(X\mathbbm{1}_p,\bar{X}\mathbbm{1}_q)$.

We follow the same spirit of \cite{Gibbons:2001wy} and take the
Euclidean metric on $\mathbb{R}^{8}$ as
\begin{equation}
\d s^2 = \d\sigma^2 + \sigma^2 \d\Omega^2_{p-1} + \d\tilde{\sigma}^2
+ \tilde{\sigma}^2 \d\Omega^2_{q-1} + \sum_{A=p+q+1}^{8} \d z^A \d
z^A\,,
\end{equation}
with $\d\Omega^2_{n}$ the round metric on the unit $n$-sphere. In
terms of these non-cartesian coordinates, the hypersurface
(\ref{hypersurface}) is explicitly given by
\begin{equation} \label{hypersurfaceexpl}
\sigma^2-(\frac{\bar{X}}{X})\tilde{\sigma}^2=\frac{R^2}{X}e^{-\tfrac{c}{2}\varphi}\sim
\tau
\end{equation}
Because the ratio $\bar{X}/X$ appears often we call it $\lambda$. If
we introduce new coordinates $r,\rho$ in the following way
\begin{equation}
\tilde{\sigma} = \rho r, \qquad \sigma = \rho(1+\lambda r^2)^{1/2}
\,,
\end{equation}
then hypersurface (\ref{hypersurfaceexpl}) is defined by $\rho^2
=\tfrac{R^2}{X}\rme^{-\tfrac{c}{2}\varphi}$ and the metric on
$\mathcal{H}^{p,q,r}$ is found to be
\begin{equation} \label{metricintspace}
\d s^2_{\mathcal{H}} =
\tfrac{R^2}{X}\rme^{-\tfrac{c}{2}\varphi}\Bigl[\frac{1+(\lambda+\lambda^2)r^2}{1+\lambda
r^2}\d r^2 +(1+\lambda r^2)\d \Omega_{p-1}^2 +r^2\d \Omega_{q-1}^2
\Bigr] + \sum_{u=p+q+1}^{8} (\d z^A)^2\,.
\end{equation}
The warp factor is
\begin{equation}
\Delta=X(1+(\lambda+\lambda^2)r^2)\,\rme^{\tfrac{c}{2}\varphi}
\end{equation}
and the 11-dimensional metric is then given by (\ref{11dmetric}).

The internal hyperbolic spaces have no non-compact isometries as
opposed to the maximally symmetric hyperboloid
\cite{Townsend:2003fx} and we cannot create a compact orbifold from
the internal hyperbolic spaces. The scaling solutions describe
internal hyperboloids with compact $\SO(p)\times\SO(q)$-symmetry and
an overall time-dependent breathing mode playing the roll of a
4-dimensional quintessence field.

\section{Discussion}

In this paper we investigated scaling solutions in $\mathcal{N}=8$
gauged supergravity. When restricted to the dilatons, the potential
becomes a sum of exponentials. We showed that, when the gauge group
is a contraction of $\SO(p,q)$, the exponentials exhibit a special
form, they have so-called affine couplings. This special form is
necessary for the existence of exact scaling solutions. We find
eternal accelerating solutions for which the barotropic fluid
vanishes. From the point of view of 11-dimensional supergravity, the
solutions correspond to timedependent geometries with non-compact
internal spaces. If we assume the presence of a barotropic fluid we
also find matter-scaling solutions. The solutions we obtained have
one running scalar and all other scalars are trapped in a saddle
point or maximum of the potential, and are therefore unstable. As
explained in \cite{Kallosh:2002gf} unstable vacua are not
necessarily a bad thing for cosmology. If a de Sitter vacuum is at a
saddle or a maximum of the potential, the universe will stop
accelerating at some point and collapses to a singularity since the
potential becomes negative. However the typical time before the
collapse is comparable to the age of our universe. We expect that a
similar statement can be made for the accelerating scaling vacua.

These dilatonic solutions describe geodesics in the scalar manifold.
We showed that geodesic scaling solutions with non-constant axions
do not exist.

Scaling solutions correspond to critical points of the cosmological
dynamical system and therefore describe the early- or late-time
behavior of general cosmological solutions. From this point of view
there is a similarity with cosmological billiards in supergravity
where the asymptotic behaviour of cosmological solutions correspond
to Kasner-type metrics and the axionic fields are constant
\cite{Fre:2005bs}\footnote{We like to thank P. Fr\'e for pointing
out this similarity.}. A more complete analysis would involve
solutions that interpolate between scaling vacua in order to
understand how the cosmic billiard behaviour is realized in gauged
extended supergravity.

In light of our findings one could wonder whether stable scaling
solutions with eternal acceleration are possible at all in
supergravity? If one lowers the amount of supersymmetry then stable
solutions are possible. In $\mathcal{N}=4$ gauged supergravity, a
(non-accelerating) stable scaling solution was found in reference
\cite{deRoo:2006ms} and, as we shortly outline below, stable eternal
accelerating scaling solutions are present in $\mathcal{N}=2$
theories. These stability properties are similar for de Sitter vacua
in supergravity, where stable vacua are only found for $\mathcal{N}
\leq 2$. So until now the only stable solutions that violate the
strong energy condition are found in $\mathcal{N}\leq 2$
supergravity.

The existence of stable scaling solutions in $\mathcal{N}=2$ gauged
supergravity follows from the fact that there exist stable de Sitter
vacua in $D=5$ \cite{Cosemans:2005sj}. If a certain supergravity has
a de Sitter solution in $4+n$ dimensions, that implies that the
system can be truncated to just gravity and a positive cosmological
constant, $\Lambda$. If we reduce this theory on an $n$-torus and
consider only one breathing mode $\varphi$ for the overall volume of
the $n$-torus,
\begin{equation}\label{metricansatz}
\d s^2= e^{\sqrt{\tfrac{n}{n+2}}\varphi}\d s_4^2 +
e^{-2\sqrt{\tfrac{1}{n(n+2)}}\varphi}\,\d \vec{z}^{\,\,2}\,,
\end{equation}
we find
\begin{equation}
S_{4}=\int \d x^{4}\sqrt{-g}_{4}[\mathcal{R}-\tfrac{1}{2}(\partial
\varphi)^2-\Lambda\,e^{\sqrt{\frac{n}{n+2}}\,\varphi} ]\,.
\end{equation}
This theory has an accelerating scaling solution
\begin{equation}
\d s_4^2=-\d \tau^2+\tau^{2\frac{n+2}{n}}\d x_3^2\,,\qquad
\sqrt{\frac{n}{n+2}}\,\varphi=-2\ln \tau + c\,.
\end{equation}
Plugging this in the metric ansatz (\ref{metricansatz}) and
redefining time via $\tilde{\tau}=\sqrt{c}\ln \tau$ we find that the
uplift of the 4-dimensional scaling solution is
\begin{equation}
\d
s_{4+n}^2=-\d\tilde{\tau}^2+e^{\frac{4}{n\sqrt{c}}\,\tilde{\tau}}\,\d
x_{3+n}^2\,.
\end{equation}
This is a 4+$n$-dimensional de Sitter universe in flat
FLRW-coordinates. When the de Sitter solution is stable, so is the
scaling solution obtained via reduction.

\section*{Acknowledgments}
We would like to thank Diederik Roest and Sven Kerstan for useful
discussions. This work is supported in part by the European
Community's Human Potential Programme under contract
MRTN-CT-2004-005104 'Constituents, fundamental forces and symmetries
of the universe' in which TVR and DBW are associated to Utrecht
University. The work of JR is supported in part by the FWO -
Vlaanderen, project G.0235.05 and by the Federal Office for
Scientific, Technical and Cultural Affairs through the
"Interuniversity Attraction Poles Programme - Belgian Science
Policy" P5/27. The work of TVR and DBW is part of the research
programme of the ``Stichting voor Fundamenteel Onderzoek der
Materie'' (FOM). JR thanks the university of Groningen for
hospitality.

\appendix
\section{Weights of $SL(8,\Real)$}\label{appendix1}
A handy basis for the weights of $SL(8,\Real)$ in the fundamental
representation is given by:
\begin{align}
& \vec{\beta}_1=(1 ,\tfrac{1}{\sqrt{3}} ,\tfrac{1}{\sqrt{6}}
,\tfrac{1}{\sqrt{10}} ,\tfrac{1}{\sqrt{15}} ,\tfrac{1}{\sqrt{21}}
,\tfrac{1}{\sqrt{28}})\\
& \vec{\beta}_2=(-1,\tfrac{1}{\sqrt{3}} ,\tfrac{1}{\sqrt{6}}
,\tfrac{1}{\sqrt{10}} ,\tfrac{1}{\sqrt{15}} ,\tfrac{1}{\sqrt{21}}
,\tfrac{1}{\sqrt{28}})\\
& \vec{\beta}_3=(0 ,\tfrac{-2}{\sqrt{3}},\tfrac{1}{\sqrt{6}}
,\tfrac{1}{\sqrt{10}} ,\tfrac{1}{\sqrt{15}} ,\tfrac{1}{\sqrt{21}}
,\tfrac{1}{\sqrt{28}})\\
& \vec{\beta}_4=(0 ,0 ,\tfrac{-3}{\sqrt{6}},\tfrac{1}{\sqrt{10}}
,\tfrac{1}{\sqrt{15}}
,\tfrac{1}{\sqrt{21}} ,\tfrac{1}{\sqrt{28}})\\
& \vec{\beta}_5=(0 ,0 ,0 ,\tfrac{-4}{\sqrt{10}},\tfrac{1}{\sqrt{15}}
,\tfrac{1}{\sqrt{21}}
,\tfrac{1}{\sqrt{28}})\\
& \vec{\beta}_6=(0 ,0 ,0 ,0
,\tfrac{-5}{\sqrt{15}},\tfrac{1}{\sqrt{21}} ,\tfrac{1}{\sqrt{28}})\\
& \vec{\beta}_7=(0 ,0 ,0 ,0
,0 ,\tfrac{-6}{\sqrt{21}},\tfrac{1}{\sqrt{28}})\\
& \vec{\beta}_8=(0 ,0 ,0 ,0 ,0 ,0 ,\tfrac{-7}{\sqrt{28}})
\end{align}

\section{Proof of lemma 5.1}

\begin{lem}
Every geodesic on the symmetric space $SL(n;\Real)/SO(n)$ can be
obtained by acting with isometries on a straight geodesic; by a
straight geodesic we mean a geodesic that has the velocity vector in
the Cartan subalgebra.
\end{lem}
\begin{proof}
Let us write $\gamma:I\subset\Real \rightarrow \mc{M}\equiv
\SL(n;\Real)/\SO(n)$ for a geodesic. Without loss of generality we
may assume that the identity point $e\cong \SO(n)$ lies in
$\gamma(I)$ since we can always multiply the geodesic by a
representative of $\gamma(t_0)^{-1}$ with $t_0\in I$ and this
geodesic goes through the origin $e$. If we can prove that by a
rotation at the origin using the compact subgroup $\SO(n)$ we can
direct any tangent vector completely in the Cartan directions, the
theorem is proved. Hence we need to know how the isotropy group acts
on the tangent space at the origin; the isotropy group keeps $e$
fixed and induces a transformation on $T\mc{M}_e$. Let $\Omega \in
\SO(n)$ generate the left-translation $x\mapsto \Omega\cdot x$.
Suppose $X$ is a vector at the origin; $X\in \mf{p}$ with $\mf{p}$
the orthogonal complement of $\mf{so}(n)$ in $\mf{sl}(n,\Real)$.
Then $X$ generates the curve $t\mapsto \e^{tX}\cdot e$, that is: $ X
= \tfrac{d}{dt}|_{0} \e^{tX}\cdot e$. The action $t_\Omega$ of
$\Omega$ on $X$ is found by:
\begin{equation}
t_\Omega(X) = \frac{\d}{\d t}|_0 \Bigl( \Omega \e^{tX}\cdot e \Bigr)
= \frac{\d}{\d t}|_0 \Bigl( \e^{\textrm{Ad}\Omega tX} \cdot e \Bigr)
= \textrm{Ad}\Omega(X)
\end{equation}
Hence we deduce that $t_\Omega(X) = \textrm{Ad}\Omega(X)$. The
adjoint representation of $\mf{sl}(n;\Real)$ decomposes under the
subgroup $\mf{so}(n)$ as follows
\begin{equation}
\mathbf{n}^2 - 1 \rightarrow \tfrac{1}{2} \mathbf{n}(\mathbf{n}-1)
\oplus \tfrac{1}{2} \mathbf{n}(\mathbf{n}+1) - 1\,.
\end{equation}
The above branching rule can be easily derived by investigating the
fundamental representation of $\mf{sl}(n;\Real)$; the subalgebra
$\mf{so}(n)$ is formed by the antisymmetric $n\times n$ matrices and
the orthogonal complement to $\mf{so}(n)$ is formed by the symmetric
traceless matrices. A symmetric traceless matrix $(m_{ij})$
transforms under the adjoint action of
$\omega_{ij}=-\omega_{ji}\in\mf{so}(n)$ as $m_{ij}\mapsto
\omega_{ik}m_{kj} + \omega_{jk}m_{ik}$, that is according to
$\tfrac{1}{2}\mathbf{n}(\mathbf{n}+1)- 1$. Hence $\mf{p}$ forms an
irreducible representation of $\mf{so}(n)$ and since every symmetric
matrix can be diagonalized using an orthogonal transformation, any
tangent vector $X\in \mf{p}$ can be rotated completely into the
Cartan subalgebra $\mf{h}\subset \mf{p}$, which we identify with the
diagonal traceless matrices in the vector representation. Hence the
lemma is proved.
\end{proof}

\bibliography{referenties3}
\bibliographystyle{utphysmodb}
\end{document}